\documentclass[aps,preprint,a4paper,tightenlines]{revtex4}%
\usepackage{amsfonts}
\usepackage{amsmath}
\usepackage{amssymb}
\usepackage{graphicx}%
\begin{document}
\title[Broad Line Approximation]{Validity of Rate Equations for Zeeman Coherences for Analysis of Nonlinear
Interaction of Atoms with Laser Radiation}
\author{Kaspars Blushs, Marcis Auzinsh}
\affiliation{Department of Physics, University of Latvia, 19 Rainis blvd, Riga, LV-1586, Latvia}
\pacs{PACS number}

\begin{abstract}
Shell document for REV\TeX{} 4.

\end{abstract}
\begin{abstract}
In this paper we, to our knowledge, for the first time obtain the rate
equations for Zeeman coherences in the broad line approximation and
steady-state balance equations directly from optical Bloch equations without
the use of the perturbation theory. The broad line approximation allows us to
use the adiabatic elimination procedure in order to eliminate the optical
coherences from the optical Bloch equations, but the steady-state condition
allows us to derive the balance equations straightforward. We compare our
approach with the perturbation theory approach as given previously and show
that our approach is more flexible in analyzing various experiments. Meanwhile
we also show the validity and limitations of the application of the rate
equations in experiments with coherent atomic excitation, when either broad
line approximation or steady-state conditions hold. Thus we have shown the
basis for modeling the coherent atomic excitation experiments by using the
relatively simple rate equations, provided that certain experimental
conditions hold.

\end{abstract}
\maketitle

\section{Introduction}

Coherent effects in laser radiation interaction with atoms and molecules play
a major role in the physics and chemistry. Applications, such as
electromagnetically induced transparency \cite{Harris1997}, laser cooling
\cite{Chu1998,CT1998,Phillips1998}, lasing without inversion \cite{Scully1989}%
, coherent population transfer \cite{berg98}, different nonlinear magneto
optical effects \cite{Budker02}, new methods for magnetometry
\cite{Scully1992,Budker2003}, coherent control of chemical reactions
\cite{Brumer2003} and many other are widely used as a powerful research tools.
Theoretical and experimental investigations of the coherent effects become
increasingly important, as they open the way for more practical applications.
Apart from some relatively simple cases, when direct solution of time
dependant Scrodinger equation can be used \cite{berg98}, usually, when one
speaks about the modelling of experiments with atomic coherent excitation, he
means the so-called "optical Bloch equations" (OBE), or Liouville equations
\cite{Allen1975,CT1977,Arimondo1996} for quantum density matrix $\rho$. These
involve both \emph{optical} and \emph{Zeeman} coherences created in an
ensemble of atoms. Zeeman coherences are quite stable and therefore it is
relatively easy to employ them practically -- for example, Zeeman coherences
are a basic ingredient of sub-Doppler and sub-recoil laser cooling mechanisms
\cite{Chu1998,CT1998,Phillips1998}.\ Optical coherences, on the other hand,
are very sensitive to a variety of factors -- collisions, finite laser
line-width, laser light fluctuations -- both in phase and in amplitude, and
many other. This means that, in describing a wide variety of atomic excitation
experiments one can neglect the optical coherences. It leads to the well known
\emph{rate equations for Zeeman coherences} \cite{CT1961,CT1977}. By saying
"rate equations" we mean that they do not couple Zeeman coherences to optical coherences.

Among the first ones to obtain the rate equations for Zeeman coherences by
neglecting the optical coherences were C.Cohen-Tannoudji and J.P.Barrat in
1961 \cite{CT1961}. They used perturbation theory to obtain the rate equations
in the so-called "broad-line approximation" (BLA) \cite{Ducloy1973}. These
rate equations were obtained by considering the excitation with light from the
spectral lamp and did not include the light induced transition effects into
the analysis . They assumed \textit{intuitively} that one can neglect the
optical coherences in case of such an excitation. No mathematical arguments
were provided, and the only justification of the used model was the good
agreement between the theory and experiment. This lack of the rigorous
mathematical argument was overcome later by C.Cohen-Tannoudji -- with a
slightly different approach, through the use of the perturbation theory and
assuming the BLA \cite{CT1977}. The BLA means, that the spectral line-width of
the laser light used in excitation of atomic transition $\triangle\omega$ is
very large compared to the natural line-width $\Gamma$ of the atomic transition%

\begin{equation}
\Delta\omega\gg\Gamma, \label{Eq1.1}%
\end{equation}
and the spacing between laser modes $\delta\omega$ is small compared to
$\Gamma$%

\begin{equation}
\delta\omega<\Gamma. \label{Eq1.2}%
\end{equation}

In this case different \textquotedblleft Bennett holes\textquotedblright,
burnt by the various modes in the Doppler profile, overlap, and the structure
caused by different holes in the atomic response disappears. If, in addition,
the modes cover all the velocity distribution, the atomic response does not
depend on the velocity of translation motion of the atom and quantum density
matrix $\rho$ refers to internal variables only. In order to use the
perturbation theory, the following condition must be satisfied:%
\begin{equation}
\triangle\omega\gg\Gamma,\Gamma_{p}, \label{Eq1.3}%
\end{equation}
where $\Gamma_{p}$ is connected to the time $T_{p}=\dfrac{1}{\Gamma_{p}}$
characterizing the evolution of density matrix $\rho$ under the effect of the
coupling with the light beam. The rate equations for Zeeman coherences,
obtained by considering the conditions (\ref{Eq1.1}) -- (\ref{Eq1.3}), are
often called the BLA equations.

In the past rate equations for Zeeman coherences in BLA were very successfully
used to analyze numerous nonlinear magneto optical effects. These include, for
example, the interaction of molecules with multimode laser radiation --
nonlinear Hanle effect, quantum beats, beat resonance, alignment to
orientation conversion in a magnetic field etc., see, for example
\cite{auz90,Auzinsh1995} and references therein. This approach from the
viewpoint of implementation in the form of computer routines seems to be
substantially less demanding technically than the OBE.

On the other hand, these rate equations for Zeeman coherences currently is not
an often used approach to describe the laser radiation interaction with atomic
gas. At first this seems obvious, as for the typical laser and atomic
line-width the BLA conditions seem to be a very special case.

However recently we have applied these equations for description of some
linear and nonlinear magneto-optical effects in stationary interaction of
alkali atoms with a broad band diode laser radiation
\cite{Alnis2001-1,Alnis2001,Alnis2003,Papoyan2002}. In these cases the BLA
clearly did not hold. Nevertheless, the agreement between simulation and
experiment was good. The detailed analysis showed that for the use of BLA
equations one does not always have to consider the BLA conditions -- a rather
striking result at a first sight. For example, in the case of a "steady-state"
excitation there actually are no limitations for the use of the BLA equations
except for the "steady-state" itself. The "steady-state" or stationary
excitation means that the excitation light does not depend on time, which
implies the same for the total density matrix $\rho(t)$ -- and this is the
case in the large number of coherent atomic excitation experiments.

What was the reason for such a good agreement between simulation and
experiment in the above experiments with alkali atoms? After a detailed
analysis it turned out, that the key factor was the fact, that the spectral
line-width of radiation from the diode lasers was mainly determined by the
phase fluctuations. The problem was, that the rigorous analysis of the
limitations of the rate equations for Zeeman coherences in case of a noisy
laser radiation seemed to be still lacking. On the other hand, there has been
a large amount of work (see the overview in \cite{Stenholm1984,Yatsenko2002})
dealing with the OBE when the exciting radiation has a finite line-width
arising from the fluctuations -- both in phase and in amplitude.

Thus in this paper we use the results obtained for the OBE and to our
knowledge for the first time obtain the rate equations for Zeeman coherences
directly from the OBE. We also compare our approach with the perturbation
theory approach \cite{CT1977} and show the advantage of our approach. We
analyze the limitations of usage of the rate equations for Zeeman coherences
in conditions of noisy laser radiation -- with an accent on analysis for
nonlinear magneto-optical effects in atoms. In the limit of large angular
momentum (molecular case) such an analysis, at least partially, was done
previously in \cite{Auz91}. In this paper our goal is to fill in this gap for
the atoms. The obtained results are in such a good agreement with experiment,
that we feel, that the relatively simple rate equations' approach (comparing
to the conventional OBE approach) is far too often undeservedly neglected,
when discussing the modelling of nonlinear magneto-optical effects in atoms.

\section{Broad band radiation interaction with atoms.}

\subsection{Exciting light.}

In our analysis of usage of rate equation for Zeeman coherences we will
describe the exciting light classically by a fluctuating electric field
$\mathbf{E}(t)$ polarized along the unit vector $\mathbf{e}$%
\begin{equation}
\mathbf{E}(t)=\varepsilon(t)\mathbf{e}+\varepsilon^{\ast}(t)\mathbf{e}^{\ast},
\label{Eq2.1}%
\end{equation}%
\begin{equation}
\varepsilon(t)=\left\vert \varepsilon_{\overline{\omega}}\right\vert
e^{-i\Phi(t)-i(\overline{\omega}-\mathbf{k}_{\overline{\omega}}\mathbf{v})t}.
\label{Eq2.2}%
\end{equation}
We account for a shift $\overline{\omega}-\mathbf{k}_{\overline{\omega}%
}\mathbf{v}$ in the laser frequency due to the Doppler effect -- $\mathbf{v}$
is the velocity of translation motion of atoms and $\mathbf{k}_{\overline
{\omega}}$ is the wave vector of the exciting light. $\overline{\omega}$ is
the center frequency of the spectrum, $\left\vert \varepsilon_{\overline
{\omega}}\right\vert $ is an amplitude of laser light field and $\Phi(t)$ is
the fluctuating phase, which gives the spectrum of the radiation a finite
bandwidth $\Delta\omega$. If the phase fluctuations are completely random,
then the line-shape of the exciting light is Lorentzian. In the case of a
laser this light corresponds to the single mode laser with randomly
fluctuating phase. In the case of a spectral lamp this light corresponds to
the lamp, where the dominant mechanism for the line-width broadening is
determined by the collisions between the radiating atoms or molecules. Note,
that for the single-mode laser the BLA condition (\ref{Eq1.2}) is not fulfilled.

The Rabi frequency $\Omega_{R}$ is determined by%
\begin{equation}
\Omega_{R}=\dfrac{d\cdot\left\vert \varepsilon_{\overline{\omega}}\right\vert
}{\hbar}, \label{Eq2.3}%
\end{equation}
where $d$ is assumed to be the strongest atomic electric dipole moment for the
transition (transitions) under consideration.

\subsection{Optical Bloch equations.}

We consider the dipole interaction of an atom with a laser field in presence
of an external static magnetic field $\mathbf{B}$. We assume that the atomic
center of mass moves classically, which means, that the only effect of the
dipole interaction of the atom with a laser field is an excitation of a
classically moving atom at the internal transitions. In this case the internal
atomic dynamics is described by the semiclassical atomic density matrix $\rho
$,\ which parametrically depends on the classical coordinates of the atomic
center of mass. We consider atoms with definite velocity $\mathbf{v}$,
illuminated by the exciting light (\ref{Eq2.1}), (\ref{Eq2.2}), resonant with
the $g\leftrightarrow e$ transition, in presence of an external static
magnetic field $\mathbf{B}$, which removes the degeneracy of the levels $g$
and $e$, so that now we consider Zeeman sublevels $g_{i}$ and $e_{i}$. In
writing OBE, see for example \cite{Stenholm1984}%
\begin{equation}
i\hbar\dfrac{\partial\rho}{\partial t}=\left[  \widetilde{H},\widetilde{\rho
}\right]  +i\hbar\widehat{R}\rho, \label{Eq3.1}%
\end{equation}
we consider only the relaxation $\widehat{R}$ due to spontaneous emission.
This means that we neglect other relaxation mechanisms, such as collisions,
fly-through relaxation etc. This assumption means, that different velocity
groups do not interact -- the density of atoms is sufficiently low. For
simplicity we also assume that the atomic transition forms a closed system --
cycling transition. In this case the spontaneous relaxation terms for a closed
system for the density matrix elements $\rho_{g_{i}g_{j}}$, $\rho_{g_{i}e_{j}%
}$, $\rho_{e_{i}g_{j}}$, $\rho_{e_{i}e_{j}}$ are:%

\begin{align}
\widehat{R}\rho_{g_{i}g_{j}}  &  =\underset{e_{i}e_{j}}{\sum}\Gamma
_{g_{i}g_{j}}^{e_{i}e_{j}}\rho_{e_{i}e_{j}},\nonumber\\
\widehat{R}\rho_{g_{i}e_{j}}  &  =-\dfrac{\Gamma}{2}\rho_{g_{i}e_{j}%
},\label{Eq3.2}\\
\widehat{R}\rho_{e_{i}g_{j}}  &  =-\dfrac{\Gamma}{2}\rho_{e_{i}g_{j}%
},\nonumber\\
\widehat{R}\rho_{e_{i}e_{j}}  &  =-\Gamma\rho_{e_{i}e_{j}},\nonumber
\end{align}
where $\Gamma_{g_{i}g_{j}}^{e_{i}e_{j}}$ describes the spontaneous relaxation
from $\rho_{e_{i}e_{j}}$ to $\rho_{g_{i}g_{j}}$ and $\Gamma$ describes the
spontaneous relaxation from $e\rightarrow g$. For the closed system it is
obvious that $\underset{g_{i}g_{j}}{\sum}\Gamma_{g_{i}g_{j}}^{e_{i}e_{j}%
}=\Gamma$. Hamiltonian $\widehat{H}=\widehat{H_{0}}+\widehat{V}$ includes the
unperturbed atomic Hamiltonian $\widehat{H_{0}}$, which depends on the
internal atomic coordinates: $\widehat{H_{0}}\left\vert \Psi_{n}\right\rangle
=E_{n}\left\vert \Psi_{n}\right\rangle $, and the dipole interaction operator
$\widehat{V}=-\widehat{\mathbf{d}}\cdot\mathbf{E}(t)$, where $\widehat
{\mathbf{d}}$\ is the electric dipole operator. Writing OBE explicitly for the
density matrix element $\rho_{ij}$, we get:%
\begin{align}
\dfrac{\partial\rho_{ij}}{\partial t}  &  =-\dfrac{i}{\hbar}\left[
\widehat{H_{0}},\rho_{ij}\right]  +\dfrac{i}{\hbar}\left[  \widehat
{\mathbf{d}}\cdot\mathbf{E}(t),\rho_{ij}\right]  +\widehat{R}\rho
_{ij}=\nonumber\\
&  =-i\omega_{ij}\rho_{ij}+\widehat{R}\rho_{ij}+\label{Eq3.3}\\
&  +\dfrac{i}{\hbar}\varepsilon\underset{k}{\sum}\left\langle i\left\vert
\mathbf{d}\cdot\mathbf{e}\right\vert k\right\rangle \rho_{kj}+\dfrac{i}{\hbar
}\varepsilon^{\ast}\underset{k}{\sum}\left\langle i\left\vert \mathbf{d}%
\cdot\mathbf{e}^{\ast}\right\vert k\right\rangle \rho_{kj}-\nonumber\\
&  -\dfrac{i}{\hbar}\varepsilon\underset{k}{\sum}\left\langle k\left\vert
\mathbf{d}\cdot\mathbf{e}\right\vert j\right\rangle \rho_{ik}-\dfrac{i}{\hbar
}\varepsilon^{\ast}\underset{k}{\sum}\left\langle k\left\vert \mathbf{d}%
\cdot\mathbf{e}^{\ast}\right\vert j\right\rangle \rho_{ik},\nonumber
\end{align}
where $\omega_{ij}=\dfrac{E_{i}-E_{j}}{\hbar}$ denotes the Zeeman splitting of
the levels $i$ and $j$. By choosing quantization axis to be parallel to the
external static magnetic field $B$, all the dependence of the density matrix
on the $B$ field is included in the splitting term $\omega_{ij}$. Thus we
arrive to the following equations for the density matrix elements $\rho
_{g_{i}g_{j}}$, $\rho_{g_{i}e_{j}}$, $\rho_{e_{i}g_{j}}$, $\rho_{e_{i}e_{j}}$:%
\begin{align}
\dfrac{\partial\rho_{g_{i}g_{j}}}{\partial t}  &  =\dfrac{i}{\hbar}%
\varepsilon\underset{e_{k}}{\sum}\left\langle g_{i}\left\vert \mathbf{d}%
\cdot\mathbf{e}\right\vert e_{k}\right\rangle \rho_{e_{k}g_{j}}+\dfrac
{i}{\hbar}\varepsilon^{\ast}\underset{e_{k}}{\sum}\left\langle g_{i}\left\vert
\mathbf{d}\cdot\mathbf{e}^{\ast}\right\vert e_{k}\right\rangle \rho
_{e_{k}g_{j}}-\nonumber\\
&  -\dfrac{i}{\hbar}\varepsilon\underset{e_{k}}{\sum}\left\langle
e_{k}\left\vert \mathbf{d}\cdot\mathbf{e}\right\vert g_{j}\right\rangle
\rho_{g_{i}e_{k}}-\dfrac{i}{\hbar}\varepsilon^{\ast}\underset{e_{k}}{\sum
}\left\langle e_{k}\left\vert \mathbf{d}\cdot\mathbf{e}^{\ast}\right\vert
g_{j}\right\rangle \rho_{g_{i}e_{k}}-i\omega_{gigj}\rho_{g_{i}g_{j}}%
+\underset{e_{i}e_{j}}{\sum}\Gamma_{g_{i}g_{j}}^{e_{i}e_{j}}\rho_{e_{i}e_{j}},
\label{Eq3.4a}%
\end{align}%
\begin{align}
\dfrac{\partial\rho_{g_{i}e_{j}}}{\partial t}  &  =\dfrac{i}{\hbar}%
\varepsilon\underset{e_{k}}{\sum}\left\langle g_{i}\left\vert \mathbf{d}%
\cdot\mathbf{e}\right\vert e_{k}\right\rangle \rho_{e_{k}e_{j}}+\dfrac
{i}{\hbar}\varepsilon^{\ast}\underset{e_{k}}{\sum}\left\langle g_{i}\left\vert
\mathbf{d}\cdot\mathbf{e}^{\ast}\right\vert e_{k}\right\rangle \rho
_{e_{k}e_{j}}-\nonumber\\
&  -\dfrac{i}{\hbar}\varepsilon\underset{g_{k}}{\sum}\left\langle
g_{k}\left\vert \mathbf{d}\cdot\mathbf{e}\right\vert e_{j}\right\rangle
\rho_{g_{i}g_{k}}-\dfrac{i}{\hbar}\varepsilon^{\ast}\underset{g_{k}}{\sum
}\left\langle g_{k}\left\vert \mathbf{d}\cdot\mathbf{e}^{\ast}\right\vert
e_{j}\right\rangle \rho_{g_{i}g_{k}}-i\omega_{giej}\rho_{g_{i}e_{j}}%
-\dfrac{\Gamma}{2}\rho_{g_{i}e_{j}}, \label{Eq3.4b}%
\end{align}%
\begin{align}
\dfrac{\partial\rho_{e_{i}g_{j}}}{\partial t}  &  =\dfrac{i}{\hbar}%
\varepsilon\underset{g_{k}}{\sum}\left\langle e_{i}\left\vert \mathbf{d}%
\cdot\mathbf{e}\right\vert g_{k}\right\rangle \rho_{g_{k}g_{j}}+\dfrac
{i}{\hbar}\varepsilon^{\ast}\underset{g_{k}}{\sum}\left\langle e_{i}\left\vert
\mathbf{d}\cdot\mathbf{e}^{\ast}\right\vert g_{k}\right\rangle \rho
_{g_{k}g_{j}}-\nonumber\\
&  -\dfrac{i}{\hbar}\varepsilon\underset{e_{k}}{\sum}\left\langle
e_{k}\left\vert \mathbf{d}\cdot\mathbf{e}\right\vert g_{j}\right\rangle
\rho_{e_{i}e_{k}}-\dfrac{i}{\hbar}\varepsilon^{\ast}\underset{e_{k}}{\sum
}\left\langle e_{k}\left\vert \mathbf{d}\cdot\mathbf{e}^{\ast}\right\vert
g_{j}\right\rangle \rho_{e_{i}e_{k}}-i\omega_{e_{i}g_{j}}\rho_{e_{i}g_{j}%
}-\dfrac{\Gamma}{2}\rho_{e_{i}g_{j}}, \label{Eq3.4c}%
\end{align}%
\begin{align}
\dfrac{\partial\rho_{e_{i}e_{j}}}{\partial t}  &  =\dfrac{i}{\hbar}%
\varepsilon\underset{g_{k}}{\sum}\left\langle e_{i}\left\vert \mathbf{d}%
\cdot\mathbf{e}\right\vert g_{k}\right\rangle \rho_{g_{k}e_{j}}+\dfrac
{i}{\hbar}\varepsilon^{\ast}\underset{g_{k}}{\sum}\left\langle e_{i}\left\vert
\mathbf{d}\cdot\mathbf{e}^{\ast}\right\vert g_{k}\right\rangle \rho
_{g_{k}e_{j}}-\nonumber\\
&  -\dfrac{i}{\hbar}\varepsilon\underset{g_{k}}{\sum}\left\langle
g_{k}\left\vert \mathbf{d}\cdot\mathbf{e}\right\vert e_{j}\right\rangle
\rho_{e_{i}g_{k}}-\dfrac{i}{\hbar}\varepsilon^{\ast}\underset{g_{k}}{\sum
}\left\langle g_{k}\left\vert \mathbf{d}\cdot\mathbf{e}^{\ast}\right\vert
e_{j}\right\rangle \rho_{e_{i}g_{k}}-i\omega_{e_{i}e_{j}}\rho_{e_{i}e_{j}%
}-\Gamma\rho_{e_{i}e_{j}}, \label{Eq3.4d}%
\end{align}

The matrix elements of the type $\left\langle e_{i}\right\vert \mathbf{d}%
\cdot\mathbf{e}\left\vert g_{j}\right\rangle $ can be calculated using the
standard angular momentum algebra \cite{Auzinsh1995,Varshalovich1988,Zare1988}.

Now, in order to eliminate the fast oscillations with optical frequency
$\overline{\omega}$, we make the following substitutions:%
\begin{align}
\rho_{g_{i}g_{j}}  &  =\widetilde{\rho_{g_{i}g_{j}}}=\rho_{g_{i}g_{j}%
},\nonumber\\
\rho_{g_{i}e_{j}}  &  =\widetilde{\rho_{g_{i}e_{j}}}e^{i(\overline{\omega
}-\mathbf{k}_{\overline{\omega}}\mathbf{v})t+i\Phi(t)},\label{Eq3.5}\\
\rho_{e_{i}g_{j}}  &  =\widetilde{\rho_{e_{i}g_{j}}}e^{-i(\overline{\omega
}-\mathbf{k}_{\overline{\omega}}\mathbf{v})t-i\Phi(t)},\nonumber\\
\rho_{e_{i}e_{j}}  &  =\widetilde{\rho_{e_{i}e_{j}}}=\rho_{e_{i}e_{j}%
}.\nonumber
\end{align}

By using the rotating wave approximation \cite{Allen1975} and neglecting terms
with double optical frequency, we arrive to:%
\begin{align}
\dfrac{\partial\rho_{g_{i}g_{j}}}{\partial t}  &  =\dfrac{i}{\hbar}\left\vert
\varepsilon_{\overline{\omega}}\right\vert \underset{e_{k}}{\sum}\left\langle
g_{i}\left\vert \mathbf{d}\cdot\mathbf{e}^{\ast}\right\vert e_{k}\right\rangle
\widetilde{\rho_{e_{k}g_{j}}}-\dfrac{i}{\hbar}\left\vert \varepsilon
_{\overline{\omega}}\right\vert \underset{e_{k}}{\sum}\left\langle
e_{k}\left\vert \mathbf{d}\cdot\mathbf{e}\right\vert g_{j}\right\rangle
\widetilde{\rho_{g_{i}e_{k}}}-\nonumber\\
&  -i\omega_{gigj}\rho_{g_{i}g_{j}}+\underset{e_{i}e_{j}}{\sum}\Gamma
_{g_{i}g_{j}}^{e_{i}e_{j}}\rho_{e_{i}e_{j}}, \label{Eq3.6a}%
\end{align}%
\begin{align}
\dfrac{\partial\widetilde{\rho_{g_{i}e_{j}}}}{\partial t}  &  =\dfrac{i}%
{\hbar}\left\vert \varepsilon_{\overline{\omega}}\right\vert \underset{e_{k}%
}{\sum}\left\langle g_{i}\left\vert \mathbf{d}\cdot\mathbf{e}^{\ast
}\right\vert e_{k}\right\rangle \rho_{e_{k}e_{j}}-\dfrac{i}{\hbar}\left\vert
\varepsilon_{\overline{\omega}}\right\vert \underset{g_{k}}{\sum}\left\langle
g_{k}\left\vert \mathbf{d}\cdot\mathbf{e}^{\ast}\right\vert e_{j}\right\rangle
\rho_{g_{i}g_{k}}-\nonumber\\
&  -i(\overline{\omega}-\mathbf{k}_{\overline{\omega}}\mathbf{v}+\omega
_{giej})\widetilde{\rho_{g_{i}e_{j}}}-\dfrac{\Gamma}{2}\widetilde{\rho
_{g_{i}e_{j}}}-i\dfrac{\partial\Phi(t)}{\partial t}\widetilde{\rho_{g_{i}%
e_{j}}}, \label{Eq3.6b}%
\end{align}%
\begin{align}
\dfrac{\partial\widetilde{\rho_{e_{i}g_{j}}}}{\partial t}  &  =\dfrac{i}%
{\hbar}\left\vert \varepsilon_{\overline{\omega}}\right\vert \underset{g_{k}%
}{\sum}\left\langle e_{i}\left\vert \mathbf{d}\cdot\mathbf{e}\right\vert
g_{k}\right\rangle \rho_{g_{k}g_{j}}-\dfrac{i}{\hbar}\left\vert \varepsilon
_{\overline{\omega}}\right\vert \underset{e_{k}}{\sum}\left\langle
e_{k}\left\vert \mathbf{d}\cdot\mathbf{e}\right\vert g_{j}\right\rangle
\rho_{e_{i}e_{k}}+\nonumber\\
&  +i(\overline{\omega}-\mathbf{k}_{\overline{\omega}}\mathbf{v}-\omega
_{e_{i}g_{j}})\widetilde{\rho_{e_{i}g_{j}}}-\dfrac{\Gamma}{2}\widetilde
{\rho_{e_{i}g_{j}}}+i\dfrac{\partial\Phi(t)}{\partial t}\widetilde{\rho
_{e_{i}g_{j}}}, \label{Eq3.6c}%
\end{align}%
\begin{align}
\dfrac{\partial\rho_{e_{i}e_{j}}}{\partial t}  &  =\dfrac{i}{\hbar}\left\vert
\varepsilon_{\overline{\omega}}\right\vert \underset{g_{k}}{\sum}\left\langle
e_{i}\left\vert \mathbf{d}\cdot\mathbf{e}\right\vert g_{k}\right\rangle
\widetilde{\rho_{g_{k}e_{j}}}-\dfrac{i}{\hbar}\left\vert \varepsilon
_{\overline{\omega}}\right\vert \underset{g_{k}}{\sum}\left\langle
g_{k}\left\vert \mathbf{d}\cdot\mathbf{e}^{\ast}\right\vert e_{j}\right\rangle
\widetilde{\rho_{e_{i}g_{k}}}-\nonumber\\
&  -i\omega_{e_{i}e_{j}}\rho_{e_{i}e_{j}}-\Gamma\rho_{e_{i}e_{j}}.
\label{Eq3.6d}%
\end{align}

\subsection{Atoms in a fluctuating optical field.}

The equations (\ref{Eq3.6a}) -- (\ref{Eq3.6d}) are stochastic differential
equations \cite{vanKampen1976} with stochastic variable $\dfrac{\partial
\Phi(t)}{\partial t}$. In an experiment, as a rule, we deal with quantities
that are averaged over the time intervals that are large in comparison with
phase fluctuation time in the excitation light source, therefore we need to
perform the statistical averaging of the above equations. In order to do that,
we solve the equations(\ref{Eq3.6b}) and (\ref{Eq3.6c}) (with initial
condition$\rho_{g_{i}e_{j}}(t_{0})=\rho_{e_{i}g_{j}}(t_{0})=0$) and then take
a formal statistical average over the fluctuating phases:%
\begin{align}
\dfrac{\partial\langle\rho_{g_{i}g_{j}}\rangle}{\partial t}  &  =\dfrac
{i}{\hbar}\left\vert \varepsilon_{\overline{\omega}}\right\vert \underset
{e_{k}}{\sum}\left\langle g_{i}\left\vert \mathbf{d}\cdot\mathbf{e}^{\ast
}\right\vert e_{k}\right\rangle \langle\widetilde{\rho_{e_{k}g_{j}}}%
\rangle-\dfrac{i}{\hbar}\left\vert \varepsilon_{\overline{\omega}}\right\vert
\underset{e_{k}}{\sum}\left\langle e_{k}\left\vert \mathbf{d}\cdot
\mathbf{e}\right\vert g_{j}\right\rangle \langle\widetilde{\rho_{g_{i}e_{k}}%
}\rangle-\nonumber\\
&  -i\omega_{gigj}\langle\rho_{g_{i}g_{j}}\rangle+\underset{e_{i}e_{j}}{\sum
}\Gamma_{g_{i}g_{j}}^{e_{i}e_{j}}\langle\rho_{e_{i}e_{j}}\rangle,
\label{Eq4.1a}%
\end{align}%
\begin{align}
\langle\widetilde{\rho_{g_{i}e_{j}}}\rangle &  =\dfrac{i}{\hbar}\left\vert
\varepsilon_{\overline{\omega}}\right\vert \underset{e_{k}}{\sum}\left\langle
g_{i}\left\vert \mathbf{d}\cdot\mathbf{e}^{\ast}\right\vert e_{k}\right\rangle
\underset{t_{0}}{\overset{t}{%
{\displaystyle\int}
}}e^{\left[  -i(\overline{\omega}-\mathbf{k}_{\overline{\omega}}%
\mathbf{v}+\omega_{g_{i}e_{j}})-\frac{\Gamma}{2}\right]  \left(  t-t^{\prime
}\right)  }\left\langle \rho_{e_{k}e_{j}}(t^{\prime})e^{-i\left[  \Phi
(t)-\Phi(t^{\prime})\right]  }\right\rangle dt^{\prime}-\nonumber\\
&  -\dfrac{i}{\hbar}\left\vert \varepsilon_{\overline{\omega}}\right\vert
\underset{g_{k}}{\sum}\left\langle g_{k}\left\vert \mathbf{d}\cdot
\mathbf{e}^{\ast}\right\vert e_{j}\right\rangle \underset{t_{0}}{\overset{t}{%
{\displaystyle\int}
}}e^{\left[  -i(\overline{\omega}-\mathbf{k}_{\overline{\omega}}%
\mathbf{v}+\omega_{g_{i}e_{j}})-\frac{\Gamma}{2}\right]  \left(  t-t^{\prime
}\right)  }\left\langle \rho_{g_{i}g_{k}}(t^{\prime})e^{-i\left[  \Phi
(t)-\Phi(t^{\prime})\right]  }\right\rangle dt^{\prime}, \label{Eq4.1b}%
\end{align}%
\begin{align}
\langle\widetilde{\rho_{e_{i}g_{j}}}\rangle &  =\dfrac{i}{\hbar}\left\vert
\varepsilon_{\overline{\omega}}\right\vert \underset{g_{k}}{\sum}\left\langle
e_{i}\left\vert \mathbf{d}\cdot\mathbf{e}\right\vert g_{k}\right\rangle
\underset{t_{0}}{\overset{t}{%
{\displaystyle\int}
}}e^{\left[  i(\overline{\omega}-\mathbf{k}_{\overline{\omega}}\mathbf{v}%
-\omega_{e_{i}g_{j}})-\frac{\Gamma}{2}\right]  \left(  t-t^{\prime}\right)
}\left\langle \rho_{g_{k}g_{j}}(t^{\prime})e^{i\left[  \Phi(t)-\Phi(t^{\prime
})\right]  }\right\rangle dt^{\prime}-\nonumber\\
&  -\dfrac{i}{\hbar}\left\vert \varepsilon_{\overline{\omega}}\right\vert
\underset{e_{k}}{\sum}\left\langle e_{k}\left\vert \mathbf{d}\cdot
\mathbf{e}\right\vert g_{j}\right\rangle \underset{t_{0}}{\overset{t}{%
{\displaystyle\int}
}}e^{\left[  i(\overline{\omega}-\mathbf{k}_{\overline{\omega}}\mathbf{v}%
-\omega_{e_{i}g_{j}})-\frac{\Gamma}{2}\right]  \left(  t-t^{\prime}\right)
}\left\langle \rho_{e_{i}e_{k}}(t^{\prime})e^{i\left[  \Phi(t)-\Phi(t^{\prime
})\right]  }\right\rangle dt^{\prime}, \label{Eq4.1c}%
\end{align}%
\begin{align}
\dfrac{\partial\left\langle \rho_{e_{i}e_{j}}\right\rangle }{\partial t}  &
=\dfrac{i}{\hbar}\left\vert \varepsilon_{\overline{\omega}}\right\vert
\underset{g_{k}}{\sum}\left\langle e_{i}\left\vert \mathbf{d}\cdot
\mathbf{e}^{\ast}\right\vert g_{k}\right\rangle \langle\widetilde{\rho
_{g_{k}e_{j}}}\rangle-\dfrac{i}{\hbar}\left\vert \varepsilon_{\overline
{\omega}}\right\vert \underset{g_{k}}{\sum}\left\langle g_{k}\left\vert
\mathbf{d}\cdot\mathbf{e}^{\ast}\right\vert e_{j}\right\rangle \langle
\widetilde{\rho_{e_{i}g_{k}}}\rangle-\nonumber\\
&  -ii\omega_{e_{i}e_{j}}\left\langle \rho_{e_{i}e_{j}}\right\rangle
-\Gamma\left\langle \rho_{e_{i}e_{j}}\right\rangle . \label{Eq4.1d}%
\end{align}

Now we employ the relation (\ref{Eq1.3}) which allows us to use the
decorrelation approximation \cite{Eberly1976,Georges1978,Georges1979}. The
decorrelation approximation means that we neglect the fluctuations of
$\rho_{a_{i}a_{j}}(t)$ ($a=e,g$) around their mean value $\left\langle
\rho_{a_{i}a_{j}}(t)\right\rangle $,\ and thus separate atom and field
variables in (\ref{Eq4.1b}) and (\ref{Eq4.1c}):%

\begin{equation}
\left\langle \rho_{a_{i}a_{j}}(t^{\prime})e^{\pm i\left[  \Phi(t)-\Phi
(t^{\prime})\right]  }\right\rangle =\left\langle \rho_{a_{i}a_{j}}(t^{\prime
})\right\rangle \left\langle e^{\pm i\left[  \Phi(t)-\Phi(t^{\prime})\right]
}\right\rangle , \label{Eq4.2}%
\end{equation}
where $a=e,g$. The decorrelation approximation in general is valid only for
Wiener-Levy-type (see below) phase fluctuations \cite{Georges1979,Georges1980}%
. In the case of a general stochastic field the decorrelation approximation
can be used as a first approximation only for weak fields below saturation
\cite{Georges1978,Georges1979}.

In order to evaluate the correlation function $\left\langle e^{\pm i\left[
\Phi(t)-\Phi(t^{\prime})\right]  }\right\rangle $, we assume two simple
models, which lead to similar results. The first one is the "phase jump" model
\cite{Burshtein1965,Burshtein1966,Eberly1984}, which assumes the phase to
remain constant, except sudden random \textquotedblleft
jumps\textquotedblright, when it changes to a new constant value. This model
is used to describe the fluctuations from the spectral lamp, when the dominant
mechanism for the line-width broadening is determined by the collisions
between the radiating atoms or molecules
\cite{Burshtein1965,Burshtein1966,Stenholm1984}. The second model is the
"phase diffusion" model \cite{Georges1979,Georges1980,Zoller1979,Dixit1980},
which assumes the continuous random diffusion of the phase. This model is used
to describe the fluctuations from the single-mode laser with fluctuating phase
\cite{Zoller1979,Dixit1980,Stenholm1984}.

\subsubsection{"Phase jump" model.}

Our analysis of phase jumps in excitation radiation is based on the detailed
analysis of the model in case of optical Bloch equations performed in
\cite{Burshtein1965,Burshtein1966,Eberly1984}. In this work we closely follow
an approach, that is analogous to the effect of instantaneous collisions on
the density matrix \cite{Stenholm1984}. The random jump\ process is Poissonian
in nature -- the probability for the phase to change $N$ times during time
$t-t^{\prime}$ is%
\begin{equation}
P_{N}=\dfrac{1}{N!}\left(  \frac{t-t^{\prime}}{T}\right)  ^{N}e^{-\frac
{t-t^{\prime}}{T}}, \label{Eq4.1.1}%
\end{equation}
where $T$ is the average time between successive phase jumps. Now we define
$\left\langle e^{\pm i\triangle\Phi}\right\rangle $ as the average phase
change during one jump. If during the time $t-t^{\prime}$ there has been only
one phase jump, then $\left\langle e^{\pm i\left[  \Phi(t)-\Phi(t^{\prime
})\right]  }\right\rangle _{1}=\left\langle e^{\pm i\triangle\Phi
}\right\rangle $. Obviously, if during the time $t-t^{\prime}$ there has been
$N$ phase jumps, then $\left\langle e^{\pm i\left[  \Phi(t)-\Phi(t^{\prime
})\right]  }\right\rangle _{N}=\left\langle e^{\pm i\triangle\Phi
}\right\rangle ^{N}$, as every jump\ in average adds one more multiplier
$\left\langle e^{\pm i\triangle\Phi}\right\rangle $. In order to get the final
expression for $\left\langle e^{\pm i\left[  \Phi(t)-\Phi(t^{\prime})\right]
}\right\rangle $, we must average over every possible number $N=0\div\infty$
of phase jumps\ during time $t-t^{\prime}$:%
\begin{align}
\left\langle e^{\pm i\left[  \Phi(t)-\Phi(t^{\prime})\right]  }\right\rangle
&  =\underset{N=0}{\overset{\infty}{\sum}}P_{N}\left\langle e^{\pm i\left[
\Phi(t)-\Phi(t^{\prime})\right]  }\right\rangle _{N}=e^{-\frac{t-t^{\prime}%
}{T}}\underset{N=0}{\overset{\infty}{\sum}}\dfrac{1}{N!}\left(  \frac
{t-t^{\prime}}{T}\left\langle e^{\pm i\triangle\Phi}\right\rangle \right)
^{N}=\nonumber\\
&  =e^{-\frac{t-t^{\prime}}{T}\left(  1-\left\langle e^{\pm i\triangle\Phi
}\right\rangle \right)  }. \label{Eq4.1.2}%
\end{align}

At this point we make further simplifications. We consider the case, when
there is no correlation of the phase values before and after the
jump\ \cite{Burshtein1965}. Then $T$ is also the correlation time of the phase
(in average after the time $T$ the phase \textquotedblleft
forgets\textquotedblright\ its past) $T=\dfrac{2}{\Delta\omega}$. We also
consider, that all the phase values occur with equal probability. Then
$\left\langle e^{\pm i\triangle\Phi}\right\rangle =0$, and (\ref{Eq4.1.2})
becomes%
\begin{equation}
\left\langle e^{\pm i\left[  \Phi(t)-\Phi(t^{\prime})\right]  }\right\rangle
=e^{-\frac{\triangle\omega}{2}\left(  t-t^{\prime}\right)  }. \label{Eq4.1.3}%
\end{equation}

This means that the spectral distribution of the exciting light is Lorentzian
with FWHM $\Delta\omega$.

\subsubsection{"Phase diffusion" model.}

Our analysis of the influence of phase diffusion of the excitation radiation
on the interaction of laser radiation with atoms we base on the phase
diffusion\ model analyzed in
\cite{Georges1978,Georges1979,Georges1980,Zoller1979,Dixit1980}. In phase
diffusion model the field has a constant amplitude, but its phase is a
fluctuating quantity, which obeys the Langevin equation%
\begin{equation}
\dfrac{d\Phi(t)}{dt}=\varsigma(t), \label{Eq4.2.1}%
\end{equation}
where $\varsigma(t)$ is a Gaussian random force with correlation function%
\begin{equation}
\left\langle \varsigma(t)\varsigma(t^{\prime})\right\rangle =b\beta
e^{-\beta\left\vert t-t^{\prime}\right\vert }, \label{Eq4.2.2}%
\end{equation}%
\begin{equation}
\left\langle \varsigma(t)\right\rangle =0, \label{Eq4.2.3}%
\end{equation}
which means, that $\varsigma(t)$ obeys the Langevin equation for Brownian
motion \cite{vanKampen1976,Stenholm1984}%
\begin{equation}
\dfrac{d}{dt}\varsigma(t)+\beta\varsigma(t)=F(t), \label{Eq4.2.4}%
\end{equation}
where $F(t)$ is a $\delta$-correlated Gaussian force fulfilling the correlation%

\begin{equation}
\left\langle F(t)F(t^{\prime})\right\rangle =2b\beta^{2}\delta(t-t^{\prime}).
\label{Eq4.2.5}%
\end{equation}

The meaning of the parameters $b$ and $\beta$ can be interpreted from the
equations (\ref{Eq4.2.1}) -- (\ref{Eq4.2.5}). $\dfrac{1}{\beta}$ is the
correlation time of the phase time derivative $\varsigma(t)$, but $b$ gives
the band-width of the field in the limit $\beta\rightarrow\infty$. Explicit
expressions for $\beta$ and $b$ in terms of fundamental laser constants are
discussed by Haken in \cite{Haken1970}, see also \cite{Avan1977}.

The spectrum of the exciting radiation, described by (\ref{Eq4.2.1}) --
(\ref{Eq4.2.5}), is given by the Fourier transform of the correlation function%
\begin{equation}
\left\langle e^{\pm i\left[  \Phi(t)-\Phi(t^{\prime})\right]  }\right\rangle
=\exp\left[  -b\left\vert t-t^{\prime}\right\vert +\dfrac{1}{\beta}%
(e^{-\beta\left\vert t-t^{\prime}\right\vert }-1\right]  . \label{Eq4.2.6}%
\end{equation}

For $\beta\gg b$ the spectrum is Lorentzian with FWHM $\triangle\omega=2b$ and
having a cut-off at frequencies $\beta$, but for $\beta\ll b$ the spectrum is
Gaussian with FWHM $\sqrt{8\ln(2)b\beta}$.

In the limit $\beta\rightarrow\infty$ the spectrum is pure Lorentzian with
FWHM $\triangle\omega=2b$ and $\varsigma(t)$ becomes $\delta$-correlated%
\begin{equation}
\left\langle \varsigma(t)\varsigma(t^{\prime})\right\rangle =2b\delta
(t-t^{\prime}), \label{Eq4.2.7}%
\end{equation}
but the phase $\Phi(t)$ obeys a Wiener-Levy stochastic process. As mentioned
above, the Wiener-Levy process is the only one, for which the decorrelation
approximation is mathematically rigorous
\cite{Georges1978,Georges1979,Georges1980}. It is easily understood, as in
this process the relevant fluctuating quantity $\varsigma(t)$ is $\delta
$-correlated (correlation time $\dfrac{1}{\beta}$ of $\varsigma(t)$ tends to
zero when $\beta\rightarrow\infty$), and thus we can always separate the
time-scales of evolution of $\left\langle \rho_{a_{i}a_{j}}(t^{\prime
})\right\rangle $ and $\left\langle e^{\pm i\left[  \Phi(t)-\Phi(t^{\prime
})\right]  }\right\rangle $ in (\ref{Eq4.2}). Wiener-Levy stochastic process
is a nonstationary Markov Gaussian process \cite{vanKampen1976}, and is
described by the Langevin equation for Brownian motion with negligible
acceleration \cite{vanKampen1976,Georges1979,Georges1980,Stenholm1984}, which
can be shown to be equivalent to the diffusion equation \cite{Stenholm1984}.
For Wiener-Levy process the relation (\ref{Eq4.2.6}) becomes:%
\begin{equation}
\left\langle e^{\pm i\left[  \Phi(t)-\Phi(t^{\prime})\right]  }\right\rangle
=\exp[-b\left\vert t-t^{\prime}\right\vert ]=e^{-\frac{\triangle\omega}%
{2}\left(  t-t^{\prime}\right)  }, \label{Eq4.2.8}%
\end{equation}
where we have used the fact, that $t\geqslant t^{\prime}$.

The Lorentz profile is not a good description of the wings of any laser
spectrum, thus this model is appropriate for rather small detunings.\ However,
as shown in\ \cite{Zoller1979,Dixit1980}, for $\beta\gg b$ (or $\beta
\gg\triangle\omega$) in the limit $\beta\gg\Gamma,\Omega_{R}$, the line-shape
of the exciting light is Lorentzian with cut-off at frequencies $\beta$. In
this case the damping term $\dfrac{\triangle\omega}{2}$ is simply multiplied
by the cut-off term, dependent\ on the detuning \cite{Zoller1979,Dixit1980}.
This corresponds to a more realistic model of the laser spectrum. We also have
to remember, that for the rotating wave approximation $\omega_{0}\gg\beta$, as
$\beta$ is the correlation time of the phase time derivative $\varsigma
(t)=\dfrac{d\Phi(t)}{dt}$.

\subsection{The effective relaxation caused by the fluctuations of the
exciting light.}

As can be seen, both approaches give similar results for time average phase
fluctuation value - the effect of the phase fluctuations on the density matrix
is simply to add the additional relaxation term, equal to the HWHM of the
exciting light. Now we use (\ref{Eq4.1.3}) and (\ref{Eq4.2.8}) to rewrite
(\ref{Eq4.1a}) - (\ref{Eq4.1d}) (for simplicity we further drop the averaging
brackets):%
\begin{align}
\dfrac{\partial\rho_{g_{i}g_{j}}}{\partial t}  &  =\dfrac{i}{\hbar}\left\vert
\varepsilon_{\overline{\omega}}\right\vert \underset{e_{k}}{\sum}\left\langle
g_{i}\left\vert \mathbf{d}\cdot\mathbf{e}^{\ast}\right\vert e_{k}\right\rangle
\widetilde{\rho_{e_{k}g_{j}}}-\dfrac{i}{\hbar}\left\vert \varepsilon
_{\overline{\omega}}\right\vert \underset{e_{k}}{\sum}\left\langle
e_{k}\left\vert \mathbf{d}\cdot\mathbf{e}\right\vert g_{j}\right\rangle
\widetilde{\rho_{g_{i}e_{k}}}-\nonumber\\
&  -i\omega_{gigj}\rho_{g_{i}g_{j}}+\underset{e_{i}e_{j}}{\sum}\Gamma
_{g_{i}g_{j}}^{e_{i}e_{j}}\rho_{e_{i}e_{j}}, \label{Eq4.3.1a}%
\end{align}%
\begin{align}
\widetilde{\rho_{g_{i}e_{j}}}  &  =\dfrac{i}{\hbar}\left\vert \varepsilon
_{\overline{\omega}}\right\vert \underset{e_{k}}{\sum}\left\langle
g_{i}\left\vert \mathbf{d}\cdot\mathbf{e}^{\ast}\right\vert e_{k}\right\rangle
\underset{t_{0}}{\overset{t}{%
{\displaystyle\int}
}}e^{\left[  -i\left(  \overline{\omega}-\mathbf{k}_{\overline{\omega}%
}\mathbf{v}+\omega_{g_{i}e_{j}}\right)  -\left(  \frac{\Gamma}{2}%
+\frac{\triangle\omega}{2}\right)  \right]  \left(  t-t^{\prime}\right)  }%
\rho_{e_{k}e_{j}}(t^{\prime})dt^{\prime}-\nonumber\\
&  -\dfrac{i}{\hbar}\left\vert \varepsilon_{\overline{\omega}}\right\vert
\underset{g_{k}}{\sum}\left\langle g_{k}\left\vert \mathbf{d}\cdot
\mathbf{e}^{\ast}\right\vert e_{j}\right\rangle \underset{t_{0}}{\overset{t}{%
{\displaystyle\int}
}}e^{\left[  -i\left(  \overline{\omega}-\mathbf{k}_{\overline{\omega}%
}\mathbf{v}+\omega_{g_{i}e_{j}}\right)  -\left(  \frac{\Gamma}{2}%
+\frac{\triangle\omega}{2}\right)  \right]  \left(  t-t^{\prime}\right)  }%
\rho_{g_{i}g_{k}}(t^{\prime})dt^{\prime}, \label{Eq4.3.1b}%
\end{align}%
\begin{align}
\widetilde{\rho_{e_{i}g_{j}}}  &  =\dfrac{i}{\hbar}\left\vert \varepsilon
_{\overline{\omega}}\right\vert \underset{g_{k}}{\sum}\left\langle
e_{i}\left\vert \mathbf{d}\cdot\mathbf{e}\right\vert g_{k}\right\rangle
\underset{t_{0}}{\overset{t}{%
{\displaystyle\int}
}}e^{\left[  i\left(  \overline{\omega}-\mathbf{k}_{\overline{\omega}%
}\mathbf{v}-\omega_{e_{i}g_{j}}\right)  -\left(  \frac{\Gamma}{2}%
+\frac{\triangle\omega}{2}\right)  \right]  \left(  t-t^{\prime}\right)  }%
\rho_{g_{k}g_{j}}(t^{\prime})dt^{\prime}-\nonumber\\
&  -\dfrac{i}{\hbar}\left\vert \varepsilon_{\overline{\omega}}\right\vert
\underset{e_{k}}{\sum}\left\langle e_{k}\left\vert \mathbf{d}\cdot
\mathbf{e}\right\vert g_{j}\right\rangle \underset{t_{0}}{\overset{t}{%
{\displaystyle\int}
}}e^{\left[  i\left(  \overline{\omega}-\mathbf{k}_{\overline{\omega}%
}\mathbf{v}-\omega_{e_{i}g_{j}}\right)  -\left(  \frac{\Gamma}{2}%
+\frac{\triangle\omega}{2}\right)  \right]  \left(  t-t^{\prime}\right)  }%
\rho_{e_{i}e_{k}}(t^{\prime})dt^{\prime}, \label{Eq4.3.1c}%
\end{align}%
\begin{align}
\dfrac{\partial\rho_{e_{i}e_{j}}}{\partial t}  &  =\dfrac{i}{\hbar}\left\vert
\varepsilon_{\overline{\omega}}\right\vert \underset{g_{k}}{\sum}\left\langle
e_{i}\left\vert \mathbf{d}\cdot\mathbf{e}\right\vert g_{k}\right\rangle
\widetilde{\rho_{g_{k}e_{j}}}-\dfrac{i}{\hbar}\left\vert \varepsilon
_{\overline{\omega}}\right\vert \underset{g_{k}}{\sum}\left\langle
g_{k}\left\vert \mathbf{d}\cdot\mathbf{e}^{\ast}\right\vert e_{j}\right\rangle
\widetilde{\rho_{e_{i}g_{k}}}-\nonumber\\
&  -i\omega_{e_{i}e_{j}}\rho_{e_{i}e_{j}}-\Gamma\rho_{e_{i}e_{j}}.
\label{Eq4.3.1d}%
\end{align}

\section{Rate equations.}

For the sake of simplicity we further assume (with $\omega_{B}$ characterizing
the Zeeman splitting):%
\begin{equation}
\triangle\omega\gg\omega_{B}, \label{Eq5.1}%
\end{equation}
though this condition can be avoided at the expense of complication of the
final rate equations. (\ref{Eq5.1}) means, that we can write:%
\begin{equation}
\left(  \frac{\Gamma}{2}+\frac{\triangle\omega}{2}\right)  +\imath\left(
\overline{\omega}-\mathbf{k}_{\overline{\omega}}\mathbf{v}+\omega
_{giej}\right)  \approx\left(  \frac{\Gamma}{2}+\frac{\triangle\omega}%
{2}\right)  +\imath\left(  \overline{\omega}-\mathbf{k}_{\overline{\omega}%
}\mathbf{v}-\omega_{0}\right)  , \label{Eq5.2}%
\end{equation}%
\begin{equation}
\left(  \frac{\Gamma}{2}+\frac{\triangle\omega}{2}\right)  -\imath\left(
\overline{\omega}-\mathbf{k}_{\overline{\omega}}\mathbf{v}-\omega_{e_{i}g_{j}%
}\right)  \approx\left(  \frac{\Gamma}{2}+\frac{\triangle\omega}{2}\right)
-\imath\left(  \overline{\omega}-\mathbf{k}_{\overline{\omega}}\mathbf{v}%
-\omega_{0}\right)  . \label{Eq5.3}%
\end{equation}

At this point we go further and assume certain conditions, which allows us to
simplify significantly the expressions for optical coherences (\ref{Eq4.3.1b})
and (\ref{Eq4.3.1c}). These conditions are either BLA (\ref{Eq1.1}) --
(\ref{Eq1.3}) or the steady-state ((\ref{Eq1.4}) - see below) conditions.
Under these circumstances the expressions for optical coherences
(\ref{Eq4.3.1b}) and (\ref{Eq4.3.1c}) become:%
\begin{equation}
\widetilde{\rho_{g_{i}e_{j}}}=\dfrac{i}{\hbar}\dfrac{\left\vert \varepsilon
_{\overline{\omega}}\right\vert }{\left(  \frac{\Gamma}{2}+\frac
{\triangle\omega}{2}\right)  +i(\overline{\omega}-\mathbf{k}_{\overline
{\omega}}\mathbf{v}-\omega_{0})}\left(  \underset{e_{k}}{\sum}\left\langle
g_{i}\left\vert \mathbf{d}\cdot\mathbf{e}^{\ast}\right\vert e_{k}\right\rangle
\rho_{e_{k}e_{j}}-\underset{g_{k}}{\sum}\left\langle g_{k}\left\vert
\mathbf{d}\cdot\mathbf{e}^{\ast}\right\vert e_{j}\right\rangle \rho
_{g_{i}g_{k}}\right)  , \label{Eq5.4a}%
\end{equation}%
\begin{equation}
\widetilde{\rho_{e_{i}g_{j}}}=\dfrac{i}{\hbar}\dfrac{\left\vert \varepsilon
_{\overline{\omega}}\right\vert }{\left(  \frac{\Gamma}{2}+\frac
{\triangle\omega}{2}\right)  -i(\overline{\omega}-\mathbf{k}_{\overline
{\omega}}\mathbf{v}-\omega_{0})}\left(  \underset{g_{k}}{\sum}\left\langle
e_{i}\left\vert \mathbf{d}\cdot\mathbf{e}\right\vert g_{k}\right\rangle
\rho_{g_{k}g_{j}}-\underset{e_{k}}{\sum}\left\langle e_{k}\left\vert
\mathbf{d}\cdot\mathbf{e}\right\vert g_{j}\right\rangle \rho_{e_{i}e_{k}%
}\right)  . \label{Eq5.4b}%
\end{equation}

Now, by substituting (\ref{Eq5.4a}) and (\ref{Eq5.4b}) in (\ref{Eq4.3.1a}) and
(\ref{Eq4.3.1d}) we arrive to the final rate equations:%
\begin{align}
\dfrac{\partial\rho_{g_{i}g_{j}}}{\partial t}  &  =\Gamma_{p}\underset
{e_{k},e_{m}}{\sum}\left\langle g_{i}\left\vert \mathbf{d}_{1}\cdot
\mathbf{e}^{\ast}\right\vert e_{k}\right\rangle \left\langle e_{m}\left\vert
\mathbf{d}_{1}\cdot\mathbf{e}\right\vert g_{j}\right\rangle \rho_{e_{k}e_{m}%
}-\nonumber\\
&  -\left(  \frac{\Gamma_{p}}{2}+i\triangle E_{p}\right)  \underset
{e_{k},g_{m}}{\sum}\left\langle g_{i}\left\vert \mathbf{d}_{1}\cdot
\mathbf{e}^{\ast}\right\vert e_{k}\right\rangle \left\langle e_{k}\left\vert
\mathbf{d}_{1}\cdot\mathbf{e}\right\vert g_{m}\right\rangle \rho_{g_{m}g_{j}%
}-\label{Eq5.5a}\\
&  -\left(  \frac{\Gamma_{p}}{2}-i\triangle E_{p}\right)  \underset
{e_{k},g_{m}}{\sum}\left\langle g_{m}\left\vert \mathbf{d}_{1}\cdot
\mathbf{e}^{\ast}\right\vert e_{k}\right\rangle \left\langle e_{k}\left\vert
\mathbf{d}_{1}\cdot\mathbf{e}\right\vert g_{j}\right\rangle \rho_{g_{i}g_{m}%
}-\nonumber\\
&  -i\omega_{gigj}\rho_{g_{i}g_{j}}+\underset{e_{i}e_{j}}{\sum}\Gamma
_{g_{i}g_{j}}^{e_{i}e_{j}}\rho_{e_{i}e_{j}},\nonumber
\end{align}%
\begin{align}
\dfrac{\partial\rho_{e_{i}e_{j}}}{\partial t}  &  =\Gamma_{p}\underset
{g_{k},g_{m}}{\sum}\left\langle e_{i}\left\vert \mathbf{d}_{1}\cdot
\mathbf{e}\right\vert g_{k}\right\rangle \left\langle g_{m}\left\vert
\mathbf{d}_{1}\cdot\mathbf{e}^{\ast}\right\vert e_{j}\right\rangle \rho
_{g_{k}g_{m}}-\nonumber\\
&  -\left(  \frac{\Gamma_{p}}{2}-i\triangle E_{p}\right)  \underset
{g_{k},e_{m}}{\sum}\left\langle e_{i}\left\vert \mathbf{d}_{1}\cdot
\mathbf{e}\right\vert g_{k}\right\rangle \left\langle g_{k}\left\vert
\mathbf{d}_{1}\cdot\mathbf{e}^{\ast}\right\vert e_{m}\right\rangle \rho
_{e_{m}e_{j}}-\label{Eq5.5b}\\
&  -\left(  \frac{\Gamma_{p}}{2}+i\triangle E_{p}\right)  \underset
{g_{k},e_{m}}{\sum}\left\langle e_{m}\left\vert \mathbf{d}_{1}\cdot
\mathbf{e}\right\vert g_{k}\right\rangle \left\langle g_{k}\left\vert
\mathbf{d}_{1}\cdot\mathbf{e}^{\ast}\right\vert e_{j}\right\rangle \rho
_{e_{i}e_{m}}-\nonumber\\
&  -i\omega_{e_{i}e_{j}}\rho_{e_{i}e_{j}}-\Gamma\rho_{e_{i}e_{j}},\nonumber
\end{align}
where $\mathbf{d}_{1}=\dfrac{\mathbf{d}}{\left\vert \mathbf{d}\right\vert }$
denotes the electric dipole moment unity vector, and thus matrix elements
$\left\langle e_{i}\right\vert \mathbf{d}\cdot\mathbf{e}\left\vert
g_{j}\right\rangle $ are written as:%
\begin{equation}
\left\langle e_{i}\right\vert \mathbf{d}\cdot\mathbf{e}\left\vert
g_{j}\right\rangle =\left\langle e_{i}\right\vert \mathbf{d}_{1}%
\cdot\mathbf{e}\left\vert g_{j}\right\rangle \left\langle e\Vert d\Vert
g\right\rangle , \label{Eq5.6}%
\end{equation}
where $\left\langle e\Vert d\Vert g\right\rangle $ is the so-called reduced
dipole matrix element. Note, that for the steady-state situation we must
consider condition (\ref{Eq1.4}) in the above equations. $\Gamma_{p}$ and
$\triangle E_{p}$ are defined as:%
\begin{equation}
\dfrac{\Gamma_{p}}{2}=\dfrac{\left\vert \varepsilon_{\overline{\omega}%
}\right\vert ^{2}}{\hbar^{2}}\times\left\vert \left\langle e\Vert d\Vert
g\right\rangle \right\vert ^{2}\times\dfrac{\left(  \frac{\Gamma}{2}%
+\frac{\triangle\omega}{2}\right)  }{\left(  \frac{\Gamma}{2}+\frac
{\triangle\omega}{2}\right)  ^{2}+(\overline{\omega}-\mathbf{k}_{\overline
{\omega}}\mathbf{v}-\omega_{0})^{2}}, \label{Eq5.7}%
\end{equation}%
\begin{equation}
\triangle E_{p}=\dfrac{\left\vert \varepsilon_{\overline{\omega}}\right\vert
^{2}}{\hbar^{2}}\times\left\vert \left\langle e\Vert d\Vert g\right\rangle
\right\vert ^{2}\times\dfrac{\overline{\omega}-\mathbf{k}_{\overline{\omega}%
}\mathbf{v}-\omega_{0}}{\left(  \frac{\Gamma}{2}+\frac{\triangle\omega}%
{2}\right)  ^{2}+(\overline{\omega}-\mathbf{k}_{\overline{\omega}}%
\mathbf{v}-\omega_{0})^{2}}. \label{Eq5.8}%
\end{equation}

$\Gamma_{p}$ is the probability per unit time of an absorption or stimulated
emission process, and $\triangle E_{p}$ describes the light shifts
\cite{CT1961} produced by the light irradiation (dynamic Stark shift). For BLA
conditions (\ref{Eq1.1}) -- (\ref{Eq1.3}) (\ref{Eq5.7}) and (\ref{Eq5.8})
become:%
\begin{equation}
\dfrac{\Gamma_{p}}{2}\approx\dfrac{\left\vert \varepsilon_{\overline{\omega}%
}\right\vert ^{2}}{\hbar^{2}}\times\left\vert \left\langle e\Vert d\Vert
g\right\rangle \right\vert ^{2}\times\dfrac{\frac{\triangle\omega}{2}}{\left(
\frac{\triangle\omega}{2}\right)  ^{2}+(\overline{\omega}-\mathbf{k}%
_{\overline{\omega}}\mathbf{v}-\omega_{0})^{2}}, \label{Eq5.9}%
\end{equation}%
\begin{equation}
\triangle E_{p}\approx\dfrac{\left\vert \varepsilon_{\overline{\omega}%
}\right\vert ^{2}}{\hbar^{2}}\times\left\vert \left\langle e\Vert d\Vert
g\right\rangle \right\vert ^{2}\times\dfrac{\overline{\omega}-\mathbf{k}%
_{\overline{\omega}}\mathbf{v}-\omega_{0}}{\left(  \frac{\triangle\omega}%
{2}\right)  ^{2}+(\overline{\omega}-\mathbf{k}_{\overline{\omega}}%
\mathbf{v}-\omega_{0})^{2}}. \label{Eq5.10}%
\end{equation}

Note also, that the phase fluctuations (described by the above models) reduce
the saturation on resonance ($\overline{\omega}-\mathbf{k}_{\overline{\omega}%
}\mathbf{v}-\omega_{0}=0$) by the factor $\dfrac{\Gamma}{\Gamma+\triangle
\omega}$, and increase the saturation far-off resonance ($\overline{\omega
}-\mathbf{k}_{\overline{\omega}}\mathbf{v}-\omega_{0}\gg\Gamma,\triangle
\omega$) by the factor $\dfrac{\Gamma+\triangle\omega}{\Gamma}$.

When the density matrix for the excited state is calculated, one can obtain
fluorescence intensity with specific polarization along the unit vector
$\mathbf{e}_{1}$\ as \cite{CT1961,Dyakonov1964,Auzinsh1995}:%
\begin{equation}
I\left(  \overrightarrow{e_{1}}\right)  =\widetilde{I_{0}}\underset
{g_{i},e_{i},e_{j}}{\sum}\left\langle e_{i}\left\vert \mathbf{d}%
\cdot\mathbf{e}_{1}^{\ast}\right\vert g_{i}\right\rangle \left\langle
g_{i}\left\vert \mathbf{d}\cdot\mathbf{e}_{1}\right\vert e_{j}\right\rangle
\rho_{e_{i}e_{j}}, \label{Eq7.1}%
\end{equation}
where $\widetilde{I_{0}}$ is a proportionality coefficient.

\section{Analysis and conclusions.}

\subsection{Perturbation theory approach.}

The obtained rate equations for Zeeman coherences coincide with equations
obtained earlier in perturbation theory approach in \cite{CT1977}. In
perturbation theory approach as small parameters are used ratios between rate
constants involved in the problem ($\Gamma_{p},\Gamma$) and line-width of the
excitation radiation $\triangle\omega$. Here we would like to stress that,
however the obtained equations coincide, the approach used in this study is
different and allows us to examine in more detail the limits of usage of rate
equations for Zeeman coherences to analyze specific experiments. To compare
both approaches, let us have a brief look in method used and conclusions
obtained with perturbation theory.

Let $T_{p}=\dfrac{1}{\Gamma_{p}}$ be the time characterizing the evolution of
density matrix $\rho$ under the effect of the coupling with the light beam. In
the following analysis it is assumed that the intensity is sufficiently low so
that $T_{p}$ is much longer than the correlation time $T=\dfrac{1}%
{\triangle\omega}$ of the light wave%
\begin{equation}
\Delta\omega\gg\Gamma_{p}. \label{Eq6.1}%
\end{equation}

Now consider a time interval $\triangle t$ such that%
\begin{equation}
T_{p},\tau\gg\Delta t\gg T, \label{Eq6.2}%
\end{equation}
where $\tau=\dfrac{1}{\Gamma}$. Since $T_{p},\tau\gg\triangle t$, one can
conclude that $\rho(t+\triangle t)-\rho(t)$ is very small and can be
calculated by perturbation theory. By using perturbation theory, it is shown
in \cite{CT1977}, that the average variation of $\rho$, $\left\langle
\rho(t+\triangle t)-\rho(t)\right\rangle $ (the average is taken over all
possible values of the random function $\varepsilon(t)$ - see below) is linear
in $\triangle t$ and only depends on $\rho(t)$%
\begin{equation}
\dfrac{\left\langle \rho(t+\triangle t)-\rho(t)\right\rangle }{\triangle
t}=\dfrac{\triangle\rho(t)}{\triangle t}. \label{Eq6.3}%
\end{equation}

This means, that we can replace $\dfrac{\triangle\rho(t)}{\triangle t}$ with
the time derivative $\dfrac{d\rho(t)}{dt}$, provided that we never use
$\dfrac{d\rho(t)}{dt}$ to describe the changes of $\rho(t)$ over time
intervals that are shorter than correlation time $T$ of the light wave, which
drives the atoms. $\dfrac{\triangle\rho(t)}{\triangle t}=\dfrac{d\rho(t)}{dt}$
is called the \textquotedblleft coarse grained\textquotedblright\ derivative
\cite{Blum1996}.

In \cite{CT1977} the exciting light is taken to be the superposition of
parallel plane waves having all the same polarization $\mathbf{e}$, but
different amplitudes $\left\vert \varepsilon_{\mu}\right\vert $, frequencies
$\omega_{\mu}$ and phases $\Phi_{\mu}$%
\begin{equation}
\mathbf{E}(t)=\varepsilon(t)\mathbf{e}+\varepsilon^{\ast}(t)\mathbf{e}^{\ast},
\label{Eq6.4}%
\end{equation}%
\begin{equation}
\varepsilon(t)=\underset{\mu}{\sum}\left\vert \varepsilon_{\mu}\right\vert
e^{-i\Phi_{\mu}-i(\omega_{\mu}-\mathbf{k}_{\overline{\omega}}\mathbf{v})t}.
\label{Eq6.5}%
\end{equation}

BLA relations (\ref{Eq1.1}) -- (\ref{Eq1.3}) hold and the relative phases of
the different modes are assumed to be completely random and thus obeying the
correlation relation:%
\begin{equation}
\left\langle e^{-i(\Phi_{\mu}-\Phi_{\mu^{\prime}})}\right\rangle =\delta
_{\mu\mu^{\prime}}, \label{Eq6.6}%
\end{equation}

The instantaneous electric field $\varepsilon(t)$ of the light wave thus may
be considered as a stationary random function, which obeys the correlation
relation:%
\begin{equation}
\left\langle \varepsilon(t)\varepsilon^{\ast}(t-\tau)\right\rangle
=\underset{\mu,\mu^{\prime}}{\sum}\left\vert \varepsilon_{\mu}\right\vert
\left\vert \varepsilon_{\mu^{\prime}}\right\vert \left\langle e^{-i(\Phi_{\mu
}-\Phi_{\mu^{\prime}})}\right\rangle e^{-i(\omega_{\mu}-\mathbf{k}%
_{\overline{\omega}}\mathbf{v})t}e^{i(\omega_{\mu^{\prime}}-\mathbf{k}%
_{\overline{\omega}}\mathbf{v})\left(  t-\tau\right)  }=\underset{\mu}{\sum
}\left\vert \varepsilon_{\mu}\right\vert ^{2}e^{-i(\omega_{\mu}-\mathbf{k}%
_{\overline{\omega}}\mathbf{v})\tau}. \label{Eq6.7}%
\end{equation}

Applying perturbation theory, after some calculations with the consideration
of (\ref{Eq6.7}) and \textquotedblleft coarse grained\textquotedblright%
\ derivative, the rate equations are obtained \cite{CT1977}, again considering
condition (\ref{Eq5.1}) for simplicity. The obtained rate equations are
exactly the same as the above derived equations (\ref{Eq5.5a}), (\ref{Eq5.5b}%
), but with $\Gamma_{p}$ and $\triangle E_{p}$ having slightly different form
as defined in (\ref{Eq5.7}) -- (\ref{Eq5.10}). This mismatch is easily
avoided, if instead of the exciting light model (\ref{Eq6.4}) -- (\ref{Eq6.7})
we take the model (\ref{Eq2.1}), (\ref{Eq2.2}), (\ref{Eq4.1.3}),
(\ref{Eq4.2.8}). Then the rate equations are the same as equations
(\ref{Eq5.5a}), (\ref{Eq5.5b}), with $\Gamma_{p}$ and $\triangle E_{p}$
defined as in (\ref{Eq5.9}) and (\ref{Eq5.10}).

Thus the perturbation theory approach is summarized as follows: we define "a
priori" the BLA conditions (\ref{Eq1.1}) -- (\ref{Eq1.3}) and then use the
perturbation theory to obtain the rate equations - and thus we are restricted
to the BLA case.

However, in our approach we obtain the "phase-averaged" OBE and then it is
possible to choose between the BLA (\ref{Eq1.1}) -- (\ref{Eq1.3}) or
steady-state (\ref{Eq1.4}) possibilities.

Thus we arrive to the conclusion stated above that the approach discussed in
this article allows to examine the limits of usage of rate equations for
Zeeman coherences to a greater detail than the perturbation theory approach.
Therefore it can be applied to analyze larger number of experimental situations.

\subsection{Velocity dependence.}

When we look at (\ref{Eq5.7}) -- (\ref{Eq5.10}), we see, that $\Gamma_{p}$
(induced transition rate) and $\triangle E_{p}$ (dynamic Stark shift) are
velocity dependent, and thus are also the equations (\ref{Eq5.5a}),
(\ref{Eq5.5b}). This means, that in describing the observable signal we need
to take into account all the velocity groups involved (note, that we have
already assumed, that different velocity groups do not interact -- the density
of atoms is sufficiently low). In standard method one has to determine the
signal dependence on velocity and then sum (integrate) over the velocities (of
course, assuming that velocity distribution is known). However, usually the
signal dependence on velocity cannot be found in analytical form, as can be
seen from (\ref{Eq5.5a}) -- (\ref{Eq5.10}). Thus a large amount of calculation
is necessary to determine this dependence -- and still it is just an approximation.

The situation is simplified only for a specific kind of experiments. For
example, if we consider a case, when exciting line-width $\triangle\omega$ is
much larger than Doppler width $\triangle\omega_{D}$ of the atomic line (as it
was originally assumed in the perturbation theory approach as given in
\cite{CT1977})%

\begin{equation}
\text{$\Delta\omega\gg$}\triangle\omega_{D}, \label{Eq8.1}%
\end{equation}
then, as mentioned above, the atomic response does not depend on the velocity
of translation motion of the atom and quantum density matrix $\rho$ refers to
internal variables only. In such a case we obtain the rate equations by simply
putting $k_{\overline{\omega}}v=0$ in (\ref{Eq5.5a}), (\ref{Eq5.5b}), which is
the same as to consider the atomic velocity group $k_{\overline{\omega}}v=0$
only. Only one velocity group is also involved in experiments with cold atomic
gases, atomic beams, etc.

However, we have successfully used the rate equations for Zeeman coherences
(\ref{Eq5.5a}), (\ref{Eq5.5b}) in modeling of various experiments
\cite{Alnis2001-1,Alnis2001,Alnis2003,Papoyan2002}. In these experiments
(\ref{Eq8.1}) clearly did not hold, nevertheless in describing experimental
signal from all velocity groups, we have used the calculated signal from just
one velocity group $k_{\overline{\omega}}v=0$ (note that $\triangle
E_{p}\left(  \mathbf{k}_{\overline{\omega}}\mathbf{v=0}\right)  =0$). It is
clear, that in this case for the experimental and simulation results to
coincide, we cannot use the exact expressions (\ref{Eq5.7}), (\ref{Eq5.9}) for
$\Gamma_{p}\left(  \mathbf{k}_{\overline{\omega}}\mathbf{v=0}\right)  $. Thus
we must consider the "effective" induced transition rate $\Gamma_{p}^{eff}$,
which in general does not coincide with $\Gamma_{p}\left(  \mathbf{k}%
_{\overline{\omega}}\mathbf{v=0}\right)  $.

Using the signal from velocity group $k_{\overline{\omega}}v=0$ as the
calculated signal is justified, if we know the relation between $\Gamma
_{p}\left(  \mathbf{k}_{\overline{\omega}}\mathbf{v=0}\right)  $ and
$\Gamma_{p}^{eff}$ in advance. In reality this relation is known only in some
specific cases -- for example, for the "steady-state" excitation with laser
intensities below saturation -- then we know, that $\Gamma_{p}^{eff}\sim
\Gamma_{p}\left(  \mathbf{k}_{\overline{\omega}}\mathbf{v=0}\right)  $, as the
signal from velocity group $k_{\overline{\omega}}v=0$ is proportional to the
signal from all velocity groups -- see \cite{Auzinsh1995}. However, in most
cases establishing the relation between $\Gamma_{p}\left(  \mathbf{k}%
_{\overline{\omega}}\mathbf{v=0}\right)  $ and $\Gamma_{p}^{eff}$ is rather
complicated, as it involves a large amount of calculations.

Therefore in analysis of experiments we have used the following approach --
the signal from velocity group $k_{\overline{\omega}}v=0$ is calculated and
then the best fit to an experiment is found -- thus experimentally finding the
relation between $\Gamma_{p}\left(  \mathbf{k}_{\overline{\omega}}%
\mathbf{v=0}\right)  $ and $\Gamma_{p}^{eff}$. In order to predict further
results, we use the extrapolation and various other mathematical techniques.
This method has proven to be successful in many cases.

\subsection{Steady-state excitation.\label{Sect1}}

As it was shown above generally, the usage of the rate equations for Zeeman
coherences for description of time dependent behavior of atoms in laser and
magnetic fields requires certain conditions regarding absorption rate
connected with light intensity and spectral width of the laser line. At the
same time very often in coherent atomic excitation experiments the
"steady-state" or stationary excitation conditions are reached -- the
excitation light does not depend on time, which implies the same for the total
density matrix $\rho(t)$. For the steady-state to happen, the system has to go
over a rather large number of cycles, which means, that the steady-state is
reached only after some time, after which it remains in this constant state
forever (unless, of course, the conditions imposed on the system are changed).
This means, that mathematically we can obtain the steady-state, if we consider
$\rho(t=\infty)$ at some certain time moment in eternity $t=\infty$, when we
can be sure, that the system has reached its steady-state - of course, if it
can reach the steady-state at all. It is also obvious, that for steady-state
the time derivative of the density matrix is zero and thus mathematically we
can also obtain the steady-state by simply putting $d\rho(t)/dt=0$ for both
optical and Zeeman coherences.

Thus under the steady-state conditions we can express the optical coherences
in terms of Zeeman coherences from the OBE straightforward, without any
assumptions. In doing so we obtain the rate equations for Zeeman coherences
$\rho_{g_{i}g_{j}}(t)$ and $\rho_{e_{i}e_{j}}(t)$, which now form a set of
linear equations, because of the steady-state condition:%
\begin{equation}
\left\{
\begin{array}
[c]{c}%
\dfrac{d\rho_{g_{i}g_{j}}(t)}{dt}=0,\\
\dfrac{d\rho_{e_{i}e_{j}}(t)}{dt}=0.
\end{array}
\right.  \label{Eq1.4}%
\end{equation}

As mentioned above, under the steady-state conditions in principle there is no
limitations in the use of the rate equations, except the steady-state
condition (\ref{Eq1.4}) itself.

\subsection{The case of large Zeeman splitting.}

In the case of large Zeeman splitting, that is, when (\ref{Eq5.1}) does not
hold, the final rate equations become more complicated. However, the
derivation procedure, of course, is still the same: we assume the already
mentioned conditions, then simplify optical coherences $\widetilde{\rho
_{g_{i}e_{j}}}$ and $\widetilde{\rho_{e_{i}g_{j}}}$ from (\ref{Eq4.3.1b}),
(\ref{Eq4.3.1c}), and substitute them in (\ref{Eq4.3.1a}), (\ref{Eq4.3.1d}).
Thus we arrive to the final rate equations, which now become more complicated
than (\ref{Eq5.5a}), (\ref{Eq5.5b}). The definitions of $\Gamma_{p}$ and
$\triangle E_{p}$ also become different than those in (\ref{Eq5.7}) --
(\ref{Eq5.10}). All the above analysis still holds.

\begin{acknowledgments}
Authors are thankful to Prof. Bruce Shore and Prof. Andrejs Reinfelds for
stimulating and enlightening discussions.
\end{acknowledgments}

\end{document}